# KGP: An R Package with Metadata from the 1000 Genomes Project

**Stephen D. Turner**

Independent Scientist

Charlottesville, Virginia

October 2, 2022

## Abstract

The 1000 Genomes Project provides sequencing data on 3,202 samples from 26 populations spanning five continental regions with no access or use restrictions. The **kgp** R package provides consistent and comprehensive metadata about samples and populations in the 1000 Genomes Project and other population sequencing data in the International Genome Sample Resource collection. The **kgp** package is distributed via the Comprehensive R Archive Network (CRAN) at `https://cran.r-project.org/package=kgp`. Source code is available on GitHub at `https://github.com/stephenturner/kgp`. Further documentation is online at `https://stephenturner.github.io/kgp/`.

***K*eywords** 1000 Genomes Project · R package · Population genetics

## 1 Introduction

The 1000 Genomes Project sequenced 2,504 individuals across 26 populations spanning five continental regions in the final Phase 3 release (Auton et al. 2015). The cohort was expanded to 3,202 individuals to include 602 trios all sequenced to 30X coverage (Byrska-Bishop et al. 2022), making the 1000 Genomes Project the largest fully open resource with whole genome sequencing data where samples are consented for public distribution without access or use restrictions (Rotimi et al. 2007). In addition to the samples from the 1000 Genomes Project, International Genome Sample Resource (IGSR) collections include over 800 samples from 186 populations collected as part of the Simons Genome Diversity Project, Human Genome Diversity Project, and Gambian Genome Variation Project (Mallick et al. 2016; Cavalli-Sforza 2005; Fairley et al. 2019).

The free and unfettered accessibility of this data has led to widespread use for method development in bioinformatics and population genetics (Zheng-Bradley and Flicek 2016), with applications including variant calling (Mills et al. 2011), imputation (Das et al. 2016), haplogrouping (Weissensteiner et al. 2016), variant prioritization (Khurana et al. 2013), population genetics (Pybus et al. 2013; Fagny et al. 2014; Enard, Messer, and Petrov 2014), and simulation for method development (Nagraj et al. 2022; Turner et al. 2022a).

All of the 1000 Genomes Project data is freely and easily accessible on the 1000 Genomes FTP site (`http://ftp.1000genomes.ebi.ac.uk/vol1/ftp/`). This includes sequence data, alignments, and variant call format (VCF) files with genotype calls for the 2,504 samples in the phase 3 release and for the 3,202 samples in the expanded high-coverage set. Metadata is also available on these samples and populations, but is spread across multiple files, requiring additional data manipulation to create a consistent and comprehensive metadata resource which may be used downstream in annotation or method development. Here I describe the **kgp** R package which provides a unified metadata and annotation resource for all 3,202 samples and all 26 populations in the 1000 Genomes Project collection.

## 2    The `kgp` package

### 2.1    Implementation

The `kgp` package is implemented as an R package with no additional external dependencies, following best practices for R data package development (Wickham 2015). Raw data is downloaded from the 1000 Genomes Project FTP site and saved to `inst/extdata` in the R package root. The commands needed to acquire these files are version controlled in a README in `inst/extdata`. A script in `data-raw` reads in each of the installed raw data files, performs necessary joins, data cleaning, and verification, following standard "tidyverse" principles (Wickham et al. 2019). Final parsed data are exposed as R data objects when the package is installed and loaded.

### 2.2    Demonstration

The `kgp` package exports four data objects. The `kgp3` object includes sample, pedigree, and population data for 2,504 samples in the Phase 3 release of the 1000 Genomes Project data. The `kgpe` object contains information for all 3,202 samples from the expanded high-coverage data. Both `kgp3` and `kgpe` contain pedigree information (family, individual, paternal, and maternal IDs), sex, population, and continental region annotations. The `kgpmeta` object contains population metadata from 26 populations across five continental regions for the 1000 Genomes Project samples in `kgp3` and `kgpe`. The `allmeta` object contains population metadata from 212 populations from the 1000 Genomes Project, Simons Genome Diversity Project, Human Genome Diversity Project, and Gambian Genome Variation Project. Both `kgpmeta` and `allmeta` contain population short codes (e.g., YRI, CEU), population names (e.g., Yoruba in Ibadan Nigeria, Utah residents with Northern and Western European ancestry); region short codes (e.g., AFR, EUR), region names (e.g., Africa, Europe), and latitude and longitude coordinates.

Table 1 shows the 1000 Genomes Project metadata available in `kgpmeta`.

Table 1: 1000 Genomes Project population metadata in `kgpmeta`.

| pop | population | reg | region | lat | lng |
|-----|-----------|-----|--------|-----|-----|
| ACB | African Caribbean in Barbados | AFR | Africa | 13.10 | -59.62 |
| ASW | African Ancestry in Southwest US | AFR | Africa | 35.48 | -97.53 |
| ESN | Esan in Nigeria | AFR | Africa | 9.07 | 7.48 |
| GWD | Gambian in Western Division, The Gambia | AFR | Africa | 13.45 | -16.58 |
| LWK | Luhya in Webuye, Kenya | AFR | Africa | -1.27 | 36.61 |
| MSL | Mende in Sierra Leone | AFR | Africa | 8.48 | -13.23 |
| YRI | Yoruba in Ibadan, Nigeria | AFR | Africa | 7.40 | 3.92 |
| CLM | Colombian in Medellin, Colombia | AMR | America | 4.58 | -74.07 |
| MXL | Mexican Ancestry in Los Angeles, California | AMR | America | 34.05 | -118.24 |
| PEL | Peruvian in Lima, Peru | AMR | America | -12.04 | -77.03 |
| PUR | Puerto Rican in Puerto Rico | AMR | America | 18.40 | -66.10 |
| CDX | Chinese Dai in Xishuangbanna, China | EAS | East Asia | 22.00 | 100.78 |
| CHB | Han Chinese in Bejing, China | EAS | East Asia | 39.92 | 116.38 |
| CHS | Southern Han Chinese, China | EAS | East Asia | 23.13 | 113.27 |
| JPT | Japanese in Tokyo, Japan | EAS | East Asia | 35.68 | 139.68 |
| KHV | Kinh in Ho Chi Minh City, Vietnam | EAS | East Asia | 10.78 | 106.68 |
| CEU | Utah residents with Northern and Western European ancestry | EUR | Europe | 40.77 | -111.89 |
| FIN | Finnish in Finland | EUR | Europe | 60.17 | 24.93 |
| GBR | British in England and Scotland | EUR | Europe | 52.49 | -1.89 |
| IBS | Iberian populations in Spain | EUR | Europe | 40.38 | -3.72 |
| TSI | Toscani in Italy | EUR | Europe | 42.10 | 12.00 |
| BEB | Bengali in Bangladesh | SAS | South Asia | 23.70 | 90.35 |
| GIH | Gujarati Indian in Houston,TX | SAS | South Asia | 29.76 | -95.37 |
| ITU | Indian Telugu in the UK | SAS | South Asia | 52.49 | -1.89 |
| PJL | Punjabi in Lahore,Pakistan | SAS | South Asia | 31.55 | 74.36 |
| STU | Sri Lankan Tamil in the UK | SAS | South Asia | 52.49 | -1.89 |

The latitude and longitude coordinates in `kgpmeta` can be used to plot a map of the locations of the 1000 Genomes populations. Not shown in Table 1 but also included in `kgpmeta` is a region color, which provides a hexadecimal color code to enable reproduction of the population data map as shown on the IGSR population data page (`https://www.internationalgenome.org/data-portal/population`). Figure 1 shows a static map produced using ggplot2, but interactive maps such as that shown on the IGSR population data portal can be created with the leaflet package.

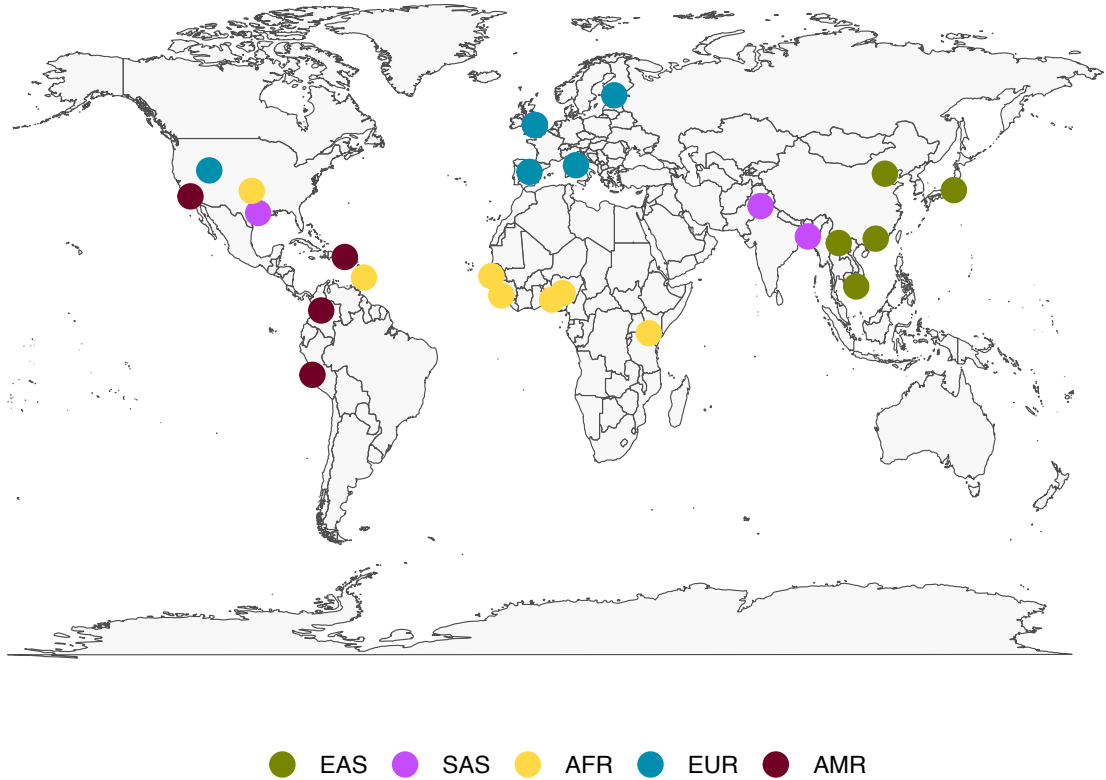

● EAS  ● SAS  ● AFR  ● EUR  ● AMR

Figure 1: Map showing locations of the 1000 Genomes Phase 3 populations.

Table 2 shows a selection of samples from `kgpe` showing pedigree information for each sample. This pedigree information could be used in downstream analysis to filter out related individuals, select only trios, or to visualize family structure using pedigree analysis packages such as kinship2 (Sinnwell, Therneau, and Schaid 2014), skater (Turner et al. 2022b), or pedtools (Vigeland 2021).

Table 2: A selection of samples from `kgpe` showing pedigree information available for each sample.

| fid | id | pid | mid | sex | pop | reg |
|---|---|---|---|---|---|---|
| BB01 | HG01881 | HG01879 | HG01880 | 2 | ACB | AFR |
| 2367 | NA19702 | NA19700 | NA19701 | 1 | ASW | AFR |
| NG06 | HG02924 | HG02923 | HG02922 | 1 | ESN | AFR |
| GB15 | HG02463 | HG02461 | HG02462 | 1 | GWD | AFR |
| SL02 | HG03056 | HG03054 | HG03055 | 1 | MSL | AFR |
| CLM03 | HG01114 | HG01112 | HG01113 | 2 | CLM | AMR |
| SH001 | HG00405 | HG00403 | HG00404 | 2 | CHS | EAS |
| VN046 | HG02015 | HG02017 | HG02016 | 1 | KHV | EAS |
| 1341 | NA06991 | NA06993 | NA06985 | 2 | CEU | EUR |
| IBS001 | HG01502 | HG01500 | HG01501 | 1 | IBS | EUR |
| BD01 | HG03008 | HG03006 | HG03007 | 1 | BEB | SAS |
| IT002 | HG03719 | HG03725 | HG03722 | 2 | ITU | SAS |

Figure 2 shows an example of a pedigree plot made by parsing the pedigree information using skater (Turner et al. 2022b) and plotting using kinship2 (Sinnwell, Therneau, and Schaid 2014). The skater package provides documentation, examples, and a vignette demonstrating how to iteratively plot all pedigrees in a given data set.

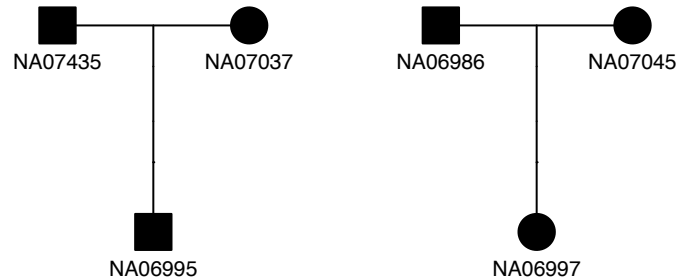

Figure 2: Trios in 1000 Genomes Project family 13291.

## 2.3 Availability

The `kgp` package is freely available as an open-source R package, available on the Comprehensive R Archive Network (CRAN) at `https://cran.r-project.org/package=kgp`.

The `kgp` package source code is available at `https://github.com/stephenturner/kgp`. Documentation is available at `https://stephenturner.github.io/kgp/` and in the R package PDF manual.

## References

Auton, Adam, Gonçalo R. Abecasis, David M. Altshuler, Richard M. Durbin, Gonçalo R. Abecasis, David R. Bentley, Aravinda Chakravarti, et al. 2015. "A Global Reference for Human Genetic Variation." *Nature* 526 (7571): 68–74. `https://doi.org/10.1038/nature15393`.

Byrska-Bishop, Marta, Uday S. Evani, Xuefang Zhao, Anna O. Basile, Haley J. Abel, Allison A. Regier, André Corvelo, et al. 2022. "High-Coverage Whole-Genome Sequencing of the Expanded 1000 Genomes Project Cohort Including 602 Trios." *Cell* 185 (18): 3426–3440.e19. `https://doi.org/10.1016/j.cell.2022.08.004`.

Cavalli-Sforza, L. Luca. 2005. "The Human Genome Diversity Project: Past, Present and Future." *Nature Reviews Genetics* 6 (4): 333–40. `https://doi.org/10.1038/nrg1596`.

Das, Sayantan, Lukas Forer, Sebastian Schönherr, Carlo Sidore, Adam E Locke, Alan Kwong, Scott I Vrieze, et al. 2016. "Next-Generation Genotype Imputation Service and Methods." *Nature Genetics* 48 (10): 1284–87. `https://doi.org/10.1038/ng.3656`.

Enard, David, Philipp W. Messer, and Dmitri A. Petrov. 2014. "Genome-Wide Signals of Positive Selection in Human Evolution." *Genome Research* 24 (6): 885–95. `https://doi.org/10.1101/gr.164822.113`.

Fagny, Maud, Etienne Patin, David Enard, Luis B. Barreiro, Lluis Quintana-Murci, and Guillaume Laval. 2014. "Exploring the Occurrence of Classic Selective Sweeps in Humans Using Whole-Genome Sequencing Data Sets." *Molecular Biology and Evolution* 31 (7): 1850–68. `https://doi.org/10.1093/molbev/msu118`.

Fairley, Susan, Ernesto Lowy-Gallego, Emily Perry, and Paul Flicek. 2019. "The International Genome Sample Resource (IGSR) Collection of Open Human Genomic Variation Resources." *Nucleic Acids Research* 48 (D1): D941–47. `https://doi.org/10.1093/nar/gkz836`.

Khurana, Ekta, Yao Fu, Vincenza Colonna, Xinmeng Jasmine Mu, Hyun Min Kang, Tuuli Lappalainen, Andrea Sboner, et al. 2013. "Integrative Annotation of Variants from 1092 Humans: Application to Cancer Genomics." *Science* 342 (6154). `https://doi.org/10.1126/science.1235587`.

Mallick, Swapan, Heng Li, Mark Lipson, Iain Mathieson, Melissa Gymrek, Fernando Racimo, Mengyao Zhao, et al. 2016. "The Simons Genome Diversity Project: 300 Genomes from 142 Diverse Populations." *Nature* 538 (7624): 201–6. `https://doi.org/10.1038/nature18964`.

Mills, Ryan E., W. Stephen Pittard, Julienne M. Mullaney, Umar Farooq, Todd H. Creasy, Anup A. Mahurkar, David M. Kemeza, et al. 2011. "Natural Genetic Variation Caused by Small Insertions and Deletions in the Human Genome." *Genome Research* 21 (6): 830–39. `https://doi.org/10.1101/gr.115907.110`.

Nagraj, V P, Matthew Scholz, Shakeel Jessa, Jianye Ge, Meng Huang, August E Woerner, Dixie Peters, Bruce Budowle, Michael D Coble, and Stephen D Turner. 2022. "Relationship Inference with Low-Coverage Whole Genome Sequencing on Forensic Samples." *Forensic Genomics* 2 (3): 81–91. `https://doi.org/10.1089/forensic.2022.0009`.

Pybus, Marc, Giovanni M. Dall'Olio, Pierre Luisi, Manu Uzkudun, Angel Carreño-Torres, Pavlos Pavlidis, Hafid Laayouni, Jaume Bertranpetit, and Johannes Engelken. 2013. "1000 Genomes Selection Browser 1.0: A Genome Browser Dedicated to Signatures of Natural Selection in Modern Humans." *Nucleic Acids Research* 42 (D1): D903–9. `https://doi.org/10.1093/nar/gkt1188`.

Rotimi, Charles, Mark Leppert, Ichiro Matsuda, Changqing Zeng, Houcan Zhang, Clement Adebamowo, Ike Ajayi, et al. 2007. "Community Engagement and Informed Consent in the International HapMap Project." *Public Health Genomics* 10 (3): 186–98. `https://doi.org/10.1159/000101761`.

Sinnwell, Jason P., Terry M. Therneau, and Daniel J. Schaid. 2014. "The Kinship2 R Package for Pedigree Data." *Human Heredity* 78 (2): 91–93. `https://doi.org/10.1159/000363105`.

Turner, Stephen D., V. P. Nagraj, Matthew Scholz, Shakeel Jessa, Carlos Acevedo, Jianye Ge, August E. Woerner, and Bruce Budowle. 2022a. "Evaluating the Impact of Dropout and Genotyping Error on SNP-Based Kinship Analysis With Forensic Samples." *Frontiers in Genetics* 13.

———. 2022b. "Skater: An R Package for SNP-Based Kinship Analysis, Testing, and Evaluation." *F1000Research* 11 (January): 18. `https://doi.org/10.12688/f1000research.76004.1`.

Vigeland, Magnus Dehli. 2021. *Pedigree Analysis in R*. Academic Press.

Weissensteiner, Hansi, Dominic Pacher, Anita Kloss-Brandstätter, Lukas Forer, Günther Specht, Hans-Jürgen Bandelt, Florian Kronenberg, Antonio Salas, and Sebastian Schönherr. 2016. "HaploGrep 2: Mitochondrial Haplogroup Classification in the Era of High-Throughput Sequencing." *Nucleic Acids Research* 44 (W1): W58–63. `https://doi.org/10.1093/nar/gkw233`.

Wickham, Hadley. 2015. *R Packages*. 1 edition. Sebastopol, CA: O'Reilly Media.

Wickham, Hadley, Mara Averick, Jennifer Bryan, Winston Chang, Lucy McGowan, Romain François, Garrett Grolemund, et al. 2019. "Welcome to the Tidyverse." *Journal of Open Source Software* 4 (43): 1686. `https://doi.org/10.21105/joss.01686`.

Zheng-Bradley, Xiangqun, and Paul Flicek. 2016. "Applications of the 1000 Genomes Project Resources." *Briefings in Functional Genomics*, July, elw027. `https://doi.org/10.1093/bfgp/elw027`.